%% file: sample-sigconf.tex
\documentclass[sigconf,natbib=true]{acmart}

\usepackage{bm}
\usepackage{enumitem}
\usepackage{multirow}
\usepackage{graphicx}
\usepackage{subfigure}
\usepackage{bbding}
\usepackage{hyperref}

\definecolor{c1}{HTML}{344C11}
\definecolor{c2}{HTML}{7e0f12}

\newcommand{\model}{\textsc{NativE}}
\newcommand{\fusionmodule}{ReDAF}
\newcommand{\advmodule}{CoMAT}

\AtBeginDocument{%
  \providecommand\BibTeX{{%
    \normalfont B\kern-0.5em{\scshape i\kern-0.25em b}\kern-0.8em\TeX}}}

\setcopyright{acmlicensed}
\copyrightyear{2024}
\acmYear{2024}
\acmDOI{XXXXXXX.XXXXXXX}

\acmConference[SIGIR 2024]{The 47th International ACM SIGIR Conference on Research and Development in Information Retrieval}{July 14-18, 2024}{Washington D.C., USA}
\acmISBN{978-1-4503-XXXX-X/18/06}

\begin{document}

\title{{\model}: Multi-modal Knowledge Graph Completion in the Wild}

\input{author_new}

\begin{abstract}
Multi-modal knowledge graph completion (MMKGC) aims to automatically discover the unobserved factual knowledge from a given multi-modal knowledge graph by collaboratively modeling the triple structure and multi-modal information from entities. However, real-world MMKGs present challenges due to their diverse and imbalanced nature, which means that the modality information can span various types (e.g., image, text, numeric, audio, video) but its distribution among entities is uneven, leading to missing modalities for certain entities. Existing works usually focus on common modalities like image and text while neglecting the imbalanced distribution phenomenon of modal information. To address these issues, we propose a comprehensive framework {\model} to achieve MMKGC in the wild. {\model} proposes a relation-guided dual adaptive fusion module that enables adaptive fusion for any modalities and employs a collaborative modality adversarial training framework to augment the imbalanced modality information. We construct a new benchmark called WildKGC with five datasets to evaluate our method. The empirical results compared with 21 recent baselines confirm the superiority of our method, consistently achieving state-of-the-art performance across different datasets and various scenarios while keeping efficient and generalizable. Our code and data are released at \textcolor{blue}{\href{https://github.com/zjukg/NATIVE}{https://github.com/zjukg/NATIVE}}.
\end{abstract}




\maketitle

\input{chapters/1-introduction}
\input{chapters/2-relatedworks}
\input{chapters/3-notation}
\input{chapters/4-method}
\input{chapters/5-experiments}
\input{chapters/6-conclusion}

\begin{acks}
This work is founded by National Natural Science Foundation of China ( NSFC62306276 / NSFCU23B2055 / NSFCU19B2027 / NSFC91846204 ), Zhejiang Provincial Natural Science Foundation of China (No. LQ23F020017), Ningbo Natural Science Foundation (2023J291), Yongjiang Talent Introduction Programme (2022A-238-G),  Fundamental Research Funds for the Central Universities (226-2023-00138).
\end{acks}

\bibliographystyle{ACM-Reference-Format}
\bibliography{sample-base}

% \appendix

\end{document}

%% file: author_new.tex
\author{Yichi Zhang}
\affiliation{
    \institution{Zhejiang University}
    \city{HangZhou}
    \state{Zhejiang}
    \country{China}
}
\email{zhangyichi2022@zju.edu.cn}

\author{Zhuo Chen}
\affiliation{
    \institution{Zhejiang University}
    \city{HangZhou}
    \state{Zhejiang}
    \country{China}
}
\email{zhuo.chen@zju.edu.cn}

\author{Lingbing Guo}
\affiliation{
    \institution{Zhejiang University}
    \city{HangZhou}
    \state{Zhejiang}
    \country{China}
}
\email{lbguo@zju.edu.cn}

\author{Yajing Xu}
\affiliation{
    \institution{Zhejiang University}
    \city{HangZhou}
    \state{Zhejiang}
    \country{China}
}
\email{yajingxu@zju.edu.cn}

\author{Binbin Hu}
\affiliation{
    \institution{Ant Group}
    \city{HangZhou}
    \state{Zhejiang}
    \country{China}
}
\email{bin.hbb@antfin.com}

\author{Ziqi Liu}
\affiliation{
    \institution{Ant Group}
    \city{HangZhou}
    \state{Zhejiang}
    \country{China}
}
\email{ziqiliu@antfin.com}

\author{Wen Zhang$^*$}
\affiliation{
    \institution{$^1$Zhejiang University\\$^2$Zhejiang University-Ant Group Joint Laboratory of Knowledge Graph\\ $^3$Alibaba-Zhejiang University Joint Institute of Frontier Technology}
    \city{HangZhou}
    \state{Zhejiang}
    \country{China}
}
\email{zhang.wen@zju.edu.cn}

\author{Huajun Chen$^*$}
\thanks{$*$ Corresponding authors}
\affiliation{
    \institution{$^1$Zhejiang University\\$^2$Zhejiang University-Ant Group Joint Laboratory of Knowledge Graph\\ $^3$Alibaba-Zhejiang University Joint Institute of Frontier Technology}
    \city{HangZhou}
    \state{Zhejiang}
    \country{China}
}
\email{huajunsir@zju.edu.cn}

%% file: chapters/1-introduction.tex
\begin{figure}[]
  \centering
\includegraphics[width=0.95\linewidth]{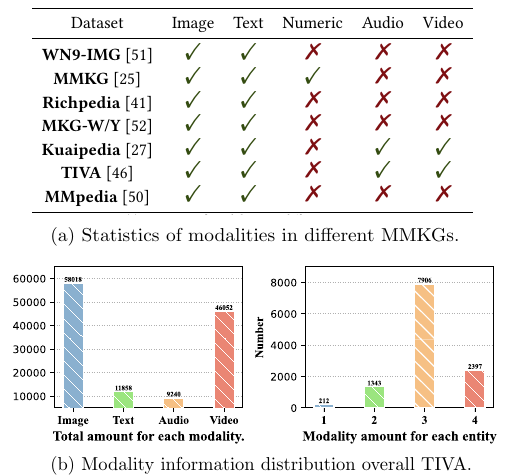}
  \vspace{-8pt}
  \caption{The diversity and imbalance nature in MMKGs. We report the modalities included in each MMKG in (a) and the statistical information about the modality information distribution across dataset/entity in TIVA in (b).}
  \label{figure::introduction}
  \vspace{-12pt}
\end{figure}

\section{Introduction}

\textbf{Multi-modal knowledge graphs (MMKGs)} \cite{DBLP:conf/esws/LiuLGNOR19-MMKG, MMKG-survey, MEAformer} represent a rich extension of traditional knowledge graphs (KGs) \cite{wang_knowledge_2017_survey1, liang_survey_2022_survey2}, enriching the structure triples in the form of \textit{(head entity, relation, tail entity)} with comprehensive multi-modal entity attributes. MMKGs are representative and have become the significant infrastructure in many AI fields such as recommender systems \cite{DBLP:conf/kdd/Wang00LC19-KGAT}, computer vision \cite{DBLP:conf/cvpr/GuCWZLW23-KGCV}, and nature language processing \cite{DBLP:conf/naacl/YasunagaRBLL21-QAGNN, DBLP:journals/corr/abs-2311-06503-knowpat, TeleBERT}.

\par Investigating how to leverage structural knowledge in KGs and MMKGs is beneficial for many areas of AI research. However, KGs and MMKGs usually face critical \textbf{incompleteness issues} as they entail many unobserved factual knowledge. This phenomenon makes knowledge graph completion (KGC) \cite{liang_survey_2022_survey2} a significant task to automatically discover new knowledge in the given KGs. Conventional KGC methods \cite{bordes_translating_2013-TransE, yang_embedding_2015-DistMult, trouillon_complex_2016-ComplEx, sun_rotate_2019-RotatE} generally emphasize learning structural embeddings to model the triple structure and measure the triple plausibility. Additionally, MMKGs accomplish the task more holistically by incorporating information from different modalities such as images and text into the KGC model, extending the task to multi-modal KGC (MMKGC) \cite{xu_relation-enhanced_2022-MMRNS, cao_otkge_2022-OTKGE, li_imf_2023-IMF}.

\par Existing MMKGC methods \cite{xie_image-embodied_2017-IKRL, sergieh_multimodal_2018-TBKGC, wang_is_2021-RSME, li_imf_2023-IMF} typically treat the multi-modal information of entities as auxiliary multi-modal embeddings and incorporate them to enhance the entity representations. However, these methods neglect two vital problems for MMKGC in real-world scenarios: the \textbf{diversity problem} and the \textbf{imbalance problem}. The diversity problem arises from the diverse modalities in contemporary information media. As the use of varied modalities such as numeric \cite{DBLP:conf/ijcai/WangISY22-KGA}, audio \cite{DBLP:conf/mm/WangMCML023-TIVA}, and video \cite{DBLP:journals/corr/abs-2211-00732-KuaiPedia} data increases in MMKG construction, a summary in Figure \ref{figure::introduction}(a) highlights the range of modalities included in the different MMKGs. However, the existing MMKGC methods are primarily designed for the prevalent image and text modalities, limiting their generalization to diverse modalities. The imbalance problem arises from the imbalanced distribution of modality information \cite{UMUEA}, implying that some important modality information would be missing in the real-world KGs. Figure \ref{figure::introduction}(b) shows the total amount and entity-wise distribution of different modalities in the TIVA \cite{DBLP:conf/mm/WangMCML023-TIVA}, confirming the uneven distribution of modalities across the datasets. Existing works do not focus on this problem or solve it by naive initialization \cite{xie_image-embodied_2017-IKRL}, leading to inadequate utilization of modality information.

\par Aiming to solve these key issues of MMKGC in the wild, we propose a novel framework {\model}, which can process and fuse \textbf{N}umeric, \textbf{A}udio, \textbf{T}ext, \textbf{I}mage, \textbf{V}ideo, and any other modalities into the \textbf{E}mbedding space with adaptive fusion and adversarial augmentation. {\model} comprises two key modules called \textbf{Relation-guided Dual Adaptive Fusion} ({\fusionmodule}) module and \textbf{Collaborative Modality Adversarial Training} ({\advmodule}) module respectively. The {\fusionmodule} facilitates diverse multi-modal fusion with any input modalities with relation guidance to moderate the weight of each modality. The {\advmodule} designs an adversarial training strategy to augment the imbalanced modality information with Wasserstein distance \cite{DBLP:journals/corr/ArjovskyCB17-WGAN} based objective. In addition, we undertake a theoretical analysis to prove the rationale of our designs. We construct a new benchmark called \textbf{WildKGC} with five MMKG datasets to evaluate our method against 21 recent baselines with further exploration. Our contributions can be summarized as three-fold:

\begin{itemize}[leftmargin=12pt]
    \item \textbf{Innovative framework}. We propose a new framework called {\model} to address the diversity and imbalance problems of MMKG in the wild. {\model} can achieve adaptive fusion with any modality with relation guidance to address the diversity problem of MMKG while augmenting the imbalanced modality information by collaborative modality adversarial training.
    \item \textbf{Theoretical analysis}. We performed a theoretical analysis to prove the legitimacy and soundness of our design.
    \item \textbf{Comprihensive experiments}. We construct a new benchmark to evaluate the MMKGC task in the wild and conduct comprehensive experiments to demonstrate the effectiveness, efficiency, and generalization of our {\model} framework.
\end{itemize}

%% file: chapters/2-relatedworks.tex
\section{Related Works}

\subsection{Knowledge Graph Completion}

\textbf{Knowledge graph completion (KGC) }\cite{liang_survey_2022_survey2} is an essential task in the community, aiming to discover the unobserved triples in the given KG. Conventional KGC methods are usually embedding-based, which embed the entities and relations of KGs into the continuous vector space and learn the embeddings based on the existing triple structure. This technique is also called \textbf{knowledge graph embedding (KGE)} \cite{wang_knowledge_2017_survey1}. Typically, conventional KGE models are designed with different score functions to measure the plausibility of triples with a general target to assign higher scores for positive triples and lower scores for negative triples. % To achieve this goal, \textbf{negative sampling (NS)} \cite{bordes_translating_2013-TransE} is widely used for model training.

\par The existing KGE models can be divided into two main categories: translation-based methods and semantic matching methods. Translation-based methods such as TransE \cite{bordes_translating_2013-TransE}, TransD \cite{ji_knowledge_2015-TransD}, RotatE \cite{sun_rotate_2019-RotatE}, OTE \cite{tang_orthogonal_2020-OTE}, and PairRE \cite{chao_pairre_2021-PairRE} modeling the triple structure as relational translation from the head entity to the tail entity, which design distance-based score functions as the plausibility measurement. Semantic matching methods such as DistMult \cite{yang_embedding_2015-DistMult}, ComplEx \cite{trouillon_complex_2016-ComplEx}, TuckER \cite{DBLP:conf/emnlp/BalazevicAH19-tucker} exploit similarity-based scoring functions based on tensor decomposition. Some methods \cite{dettmers_convolutional_2018-ConvE,DBLP:conf/nips/ZhuZXT21-NBFNET, DBLP:journals/corr/abs-2310-06671-KOPA,DBLP:journals/corr/KG-BERT,KGC1,KGC2} also attempt to extract structural semantics with deep neural networks. 
% There are also some fine-tuning-based KGC methods \cite{DBLP:journals/corr/KG-BERT, DBLP:conf/coling/MTL-KGC, DBLP:conf/www/STAR,DBLP:conf/acl/KGT5, DBLP:conf/coling/KGS2S, DBLP:conf/acl/PKGC, DBLP:journals/corr/kgllama, DBLP:journals/corr/abs-2310-06671-KOPA} which utilize the pre-trained language models \cite{DBLP:conf/nips/transformer} like BERT \cite{DBLP:conf/naacl/DevlinCLT19_BERT} and T5 \cite{DBLP:journals/jmlr/T5} to accomplish the KGC tasks as text classification or generation in the contextualized settings rather than learning explicit embeddings.

\subsection{Multi-modal Knowledge Graph Completion}

\textbf{Multi-modal knowledge graph completion (MMKGC)} further considers utilizing the complex multi-modal information in the MMKGs to benefit the KGC. Current mainstream methods usually extend the conventional KGE models with more flexible multi-modal embeddings of entities such as visual embeddings and textual embeddings. These embeddings are extracted with pre-trained models and represent the entity feature from multi-views.  In our taxonomy, the MMKGC methods can be further divided into three categories. The first category is the modal fusion methods. These methods \cite{xie_image-embodied_2017-IKRL, sergieh_multimodal_2018-TBKGC, pezeshkpour_embedding_2018-MKBE, wang_multimodal_2019-TransAE, wang_is_2021-RSME, cao_otkge_2022-OTKGE, lee_vista_2023-VISTA,AdaMF-MAT, SNAG} design elegant approaches to achieve multi-modal fusion in the same representation space.
The second category is modal ensemble methods. These methods \cite{zhao_mose_2022-MOSE, li_imf_2023-IMF} learn the respective models for each modality and make joint predictions with ensemble learning.
The third category is the negative sampling (NS) enhanced methods. As mentioned before, NS is an important technology for KGE model training. Therefore, some methods \cite{DBLP:conf/ijcnn/ZhangCZ23-MANS, xu_relation-enhanced_2022-MMRNS, DBLP:journals/corr/abs-2209-07084-VBKGC} attempt to enhance the NS process and generate high-quality negative samples by utilizing the multi-modal information. The problem scenarios of existing MMKGC methods are relatively simple and usually consider the text and image modalities, while some of the work just considers the numerical modality. Also, they do not consider the data imbalance problem in MMKGs. In our work, we plan to accomplish the MMKGC task with a more unified perspective for more modalities and more complex multi-modal data distribution in the wild.

\subsection{Generative Adversarial Networks}
Generative adversarial networks (GAN) \cite{goodfellow_generative_2014_gan} is a milestone progress in the field of deep learning. GAN proposes to train a pair of discriminator and generator by playing a min-max game between them and achieve better performance, which has been widely used in various fields such as computer vision \cite{DBLP:journals/corr/ArjovskyCB17-WGAN, DBLP:conf/cvpr/KarrasLA19-styleGAN}, natural language processing \cite{DBLP:conf/aaai/YuZWY17-GAN4NLP, DBLP:conf/acl/CroceCB20-GAN4NLP2}, information retrieval \cite{DBLP:conf/sigir/WangYZGXWZZ17-IRGAN, DBLP:conf/iclr/ZhangGS0DC22-GAN4IR}, and recommender systems \cite{DBLP:conf/www/WeiHXZ23-MMSSL, DBLP:conf/aaai/YuZWY17-GAN4NLP, DBLP:conf/www/WangX000C20-GAN4Rec2}. In the KGC field, there are some methods \cite{cai_kbgan_2018-KBGAN, wang_incorporating_2018-IGAN, tang_positive-unlabeled_2022-PUDA, DBLP:journals/apin/LuWJHL22-MMKRL} employs a GAN-based framework to enhance the negative sampling \cite{bordes_translating_2013-TransE} process. For example, KBGAN \cite{cai_kbgan_2018-KBGAN} utilizes reinforcement learning (RL) with GAN to learn a better sampling policy. MMKRL \cite{DBLP:journals/apin/LuWJHL22-MMKRL} designs an adversarial training strategy for MMKGC. However, these methods tend to be oriented towards conventional KGC and often require RL, which leads to a more limited effect. Our work is the first to involve multi-modal entity information in MMKGCs and propose a unified framework to enhance MMKGC models.

%% file: chapters/3-notation.tex
\section{Task Definition}
\label{section3}
\input{chapters/4-model}
A KG can be typically represented as $\mathcal{K}=(\mathcal{E}, \mathcal{R}, \mathcal{T})$ where $\mathcal{E}, \mathcal{R}$ are the entity set, the relation set respectively. $\mathcal{T}=\{(h, r, t)\mid h, t \in\mathcal{E}, r\in\mathcal{R}\}$ is the triple set. Furthermore, MMKGs have a modality set denoted as $\mathcal{M}$, encapsulating different modalities (image $I$, text $T$, numeric $N$, audio $A$, video $V$) in the MMKGs. For a given modality $m\in\mathcal{M}$, the set of modal information is denoted as $\mathcal{X}_m$. For an entity $e\in\mathcal{E}$, its modality information $m$ is denoted as $\mathcal{X}_m(e)$, which is an empty set $\empty$ if the corresponding modal information is missing. For different modalities, the elements in it have different forms. For instance, $\mathcal{X}_m(e)$ can be a set of images when $m=I$ and some video clips when $m=V$. Note that the graph structure ($S$) is also an intrinsic modality for each entity and the structural information is already embodied in the triple set $\mathcal{T}$.

\par The general purpose of the MMKGC task is to learn a score function $\mathcal{F}: \mathcal{E}\times\mathcal{R}\times\mathcal{E}\rightarrow \mathbb{R}$ to discriminate the plausibility of a given triple $(h, r, t)$. In this context, a higher score implies a more plausible triple. To enable differentiable computations with gradient-based optimization, entities and relations are embedded into continuous vector spaces, which is called knowledge graph embedding (KGE). The embeddings of entity $e$ and relation $r$ are denoted as $\bm{e}\in \mathbb{R}^{d_e}$ and $\bm{r}\in \mathbb{R}^{d_r}$, where $d_e, d_r$ are the embedding dimensions. Convectional KGE models focus solely on the structural information of triple, rendering these embeddings as \textbf{structural embeddings}. The score function $\mathcal{F}(h, r, t)$ leverages the structural embeddings to calculate the triple scores. For MMKGs, the multi-modal entity information should also be considered in the score function, implying additional embeddings $\bm{e}_{m}$ for varying modalities $m\in\mathcal{M}$ to represent the multi-modal feature of an entity $e$. This expansion consequently complicates the score function by considering multi-modal embedding integration. During the training stage, negative sampling (NS) \cite{bordes_translating_2013-TransE} is widely used to construct manual negative triples for contrastive learning as KGs usually consist of only observed positive triples. Given a positive triple $(h, r, t)$, the head $h$ or tail $t$ is randomly replaced by another entity $e\in\mathcal{E}$ in the NS process. The negative triple set can be denoted as $\mathcal{T}'=\{(h',r,t)\mid (h, r, t)\in\mathcal{T}\cap h'\in\mathcal{E}\setminus\{h\}\} \cup \{(h,r,t')\mid (h, r, t)\in\mathcal{T}\cap t'\in\mathcal{E}\setminus\{t\}\}$.
During the inference stage, the KGC model is usually evaluated with the link prediction task \cite{bordes_translating_2013-TransE}. The target of link prediction is to predict the missing head or tail entity in the given query $(?, r, t)$ or $(h, r, ?)$. For instance, in tail prediction, the entire set of entities $\mathcal{E}$ will be the candidate set during evaluation. For each $e\in\mathcal{E}$, the plausibility score of the triple $(h, r, e)$ is calculated and then ranked across the entire candidate set. A higher rank of the ground truth $(h, r, t)$ represents better model performance.

%% file: chapters/4-model.tex
\begin{figure*}[]
  \centering
\includegraphics[width=0.9\linewidth]{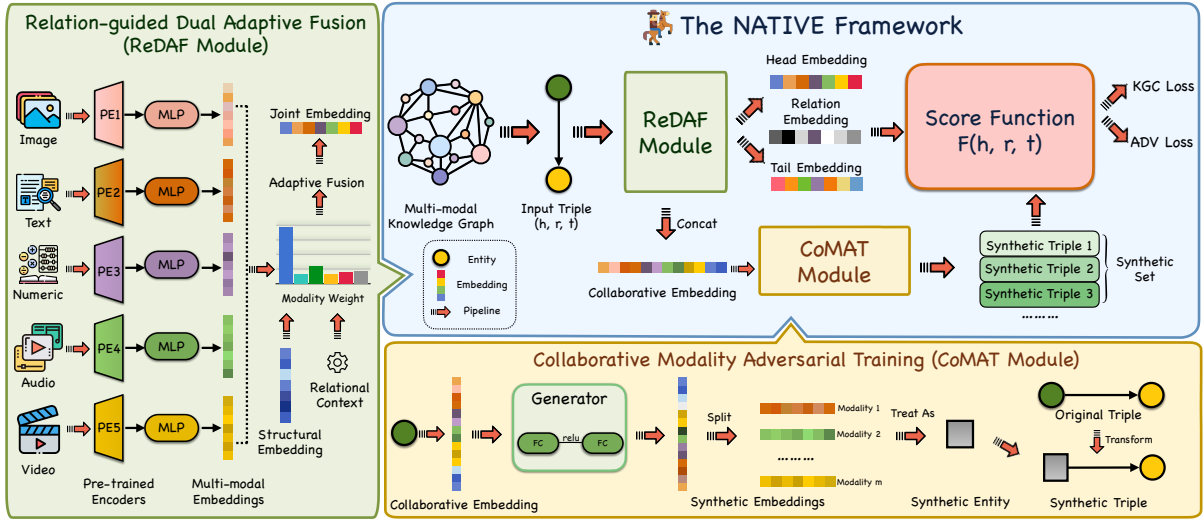}
  \vspace{-8pt}
  \caption{The overview of our {\model} framework. {\model} consists of two main modules called relation-guided dual adaptive fusion ({\fusionmodule}) module and collaborative modality adversarial training ({\advmodule}) module respectively. {\fusionmodule} is designed to fuse any input modality with modality adaptive weights and relational guidance. {\advmodule} aims to augment the imbalanced modality information in an adversarial manner by constructing synthetic triples to play a min-max game.}
  \label{figure::model}
  \vspace{-12pt}
\end{figure*}

%% file: chapters/4-method.tex
\section{Methodology}

In this section, we will introduce the proposed MMKGC framework in detail. We refer to our model as {\model}, designed to represent and combine multiple data modalities (\textbf{N}umeric, \textbf{A}udio, \textbf{T}ext, \textbf{I}mage, \textbf{V}ideo, and more) from MMKGs into multi-modal \textbf{E}mbeddings with adversarial augmentation. This ability readies {\model} to deliver robust prediction capabilities in the wild while facing diversity and imbalance problems. In {\model}, we design two new modules called relation-guided dual adaptive fusion and collaborative modality adversarial training to address the problems mentioned before.

\subsection{Modality Encoding}

To leverage the modality information, we first perform modality encoding to capture modality feature, which is a very common step in different MMKGC methods. The raw feature $f_m$ of the entity $e$ in the modality $m$ is extracted as:
\begin{equation}
    f_{m}=\frac{1}{|\mathcal{X}_m(e)|}\sum\nolimits_{x_{e, m}\in \mathcal{X}_m(e)} \mathbf{PE}_{m}(x_{e, m})
\end{equation}
where $x_{e, m}$ is one of the modality $m$ elements for entity $e$ and $\mathbf{PE}_{m}$ is the corresponding pre-trained encoder. The encoders, typically pre-trained on extensive datasets, extract deep semantic characteristics distinctive to each modality. For example, we can employ BERT \cite{DBLP:conf/naacl/DevlinCLT19_BERT} as the textual encoder and VGG16 \cite{DBLP:journals/corr/SimonyanZ14a_VGG} as the image encoder. The numeric information can be also extended as a sequence and employ BERT to capture its modality information. The pre-trained encoders are frozen to capture modality features and will not be fine-tuned during the training process. Moreover, the modal embedding for entity $e\in\mathcal{E}$ of modality $m\in\mathcal{M}$ is indicated as $\bm{e}_{m}=\mathcal{P}_{m}\left(f_{m}\right) \in \mathbb{R}^{d_e}$,
where ${d_e}$ is the embedding dimension and $\mathcal{P}_m$ is a projection layer for the modality $m$, aiming to project the different modal embeddings into the same vector space. Each $\mathcal{P}_m$ comprises a two-layer MLP with ReLU \cite{DBLP:journals/jmlr/GlorotBB11_relu} as the activation function.

\par The modality encoding process yields distinct modal embeddings $\bm{e}_m$ for each entity $e$. In the MMKG context, embeddings from other modalities serve as auxiliary information to enhance the structure of the entity. For uniform representation of multi-modal information, we assign the structural modality as $S$ and the structural embeddings $\bm{e}$ as $\bm{e}_S$. However, structural embeddings for entities deviate from other modalities. They are set to be \textbf{learnable parameters} to fit the triple structural information rather than relying on the modality encoding offered by pre-trained encoders.

\subsection{Relation-guided Dual Adaptive Fusion}

\par To extract the rich semantic information contained within entities, it is customary to fuse these multi-modal embeddings. Existing methods often employ fusion mechanisms such as concatenation \cite{sergieh_multimodal_2018-TBKGC}, dot product \cite{li_imf_2023-IMF}, or gating \cite{wang_is_2021-RSME} to accomplish modality fusion. Nevertheless, these methods are primarily designed for specific modalities such as text and images, which may not adequately serve scenarios intertwined with a richer mix of modalities. Furthermore, given the inherent imbalance in KGs, different modalities need to play different roles and offer diverse evidence for robust prediction within varying contexts.

\par To address the mentioned problems, we propose a \textbf{Relation-guided Dual Adaptive Fusion} (abbreviated as \textbf{{\fusionmodule}}) module within our framework. {\fusionmodule} includes an adaptive fusion mechanism using a set of adaptive weights $\omega_m$ for different modalities, which can be dynamically adjusted when the modal information is imbalanced. For example, when a certain modal is missing and the corresponding modal embedding is randomly initialized, the adaptive weight will be decreased for this modality. Meanwhile, we design a \textbf{relational-wise temperature} $\zeta_r$ to further modulate the weight distribution, thereby providing relational context for entity representation. This leads to dynamic weight adjustments populated both over different entities and under different relational contexts, which is the reason we name this module dual adaptive fusion. In more specific terms, the head entity $h$ of the triple $(h, r, t)$ has a relational context $r$ and the adaptive weight for each modal $m$ of $h$ is denoted as:

\begin{equation}
    \omega_m(h, r)=\frac{\exp(\mathcal{V}\odot \mathrm{Tanh}(\bm{h}_{m})/\sigma(\zeta_r))}{\sum_{n\in\mathcal{M}\cup\{S\}}\exp(\mathcal{V}\odot \mathrm{Tanh}(\bm{h}_{n})/\sigma(\zeta_r))}
\end{equation}
where $\mathcal{V}$ is a learnable vector and $\odot$ is the point-wise operator. $\mathrm{Tanh}()$ is the tanh function. $\sigma$ represents the sigmoid function to limit the relational-wise temperature in $(0, 1)$, aiming to amplify the differences between different modal weights.  With the adaptive weights, the joint embedding of the head entity $h$ is aggregated as:

\begin{equation}
    \bm{h}_{joint}=\sum\nolimits_{m\in M\cup\{S\}}\omega_m(h, r)\bm{h}_{m}
\end{equation}
The joint embedding $\bm{t}_{joint}$ of tail $t$ can be also obtained similarly. 

\par Classical MMKGC methods typically employ entity-centric modality fusion methods, regardless of their triple context. The strength of our design lies in its ability to dynamically modulate the modality weights for different entities across different triples, permitting these weights to be dynamically adjusted by their relational context. Therefore, {\fusionmodule} facilitates dual-adaptive multi-modal fusion, considering both the modal information of the entity and the relational context. Furthermore, {\fusionmodule} is a versatile modality fusion module, capable of processing \textbf{an unlimited number of input modalities} rather than considering specified modalities.

\par Once obtaining the joint embedding of entities, we employ a RotatE \cite{sun_rotate_2019-RotatE} score function to discriminate the triple plausibility, the score function is denoted as:
\begin{equation}
\label{score}
    \mathcal{F}(h, r, t)=-|| \bm{h}_{joint}\circ\bm{r}-\bm{t}_{joint}||
\end{equation}
where $\circ$ is the rotation operator in complex space. A higher score represents higher triple plausibility. We chose RotatE as the score function as RotatE can model the most common relational patterns. During training, we employ a negative sampling-based loss function to optimize the parameters, which can be represented as:

\begin{equation}
    \begin{aligned}
        \mathcal{L}_{kgc}=&\sum\nolimits_{(h, r, t)\in \mathcal{T}}-\log\sigma(\gamma+\mathcal{F}(h,r,t))\\
        -&\sum_{i=1}^{K}p(h_{i}',r_{i}',t_{i}')\log\sigma(-\mathcal{F}(h_{i}',r_{i}',t_{i}')-\gamma)
    \end{aligned}
\end{equation}
where $\sigma$ is the sigmoid function; $\gamma$ is a fixed margin; $(h_{i}',r_{i}',t_{i}')\in\mathcal{T}', (i=1,2,\dots, K)$ are $K$ negative samples for triple $(h, r, t)$. Besides, $p(h_{i}',r_{i}',t_{i}')$ is the self-adversarial weight \cite{sun_rotate_2019-RotatE} proposed in RotatE, it can be denoted as:

\begin{equation}
p(h_{i}',r_{i}',t_{i}')=\frac{\exp\left(\beta\mathcal{F}(h_{i}',r_{i}',t_{i}')\right)}{\sum_{j=1}^k\exp\left(\beta\mathcal{F}(h_{j}',r_{j}',t_{j}')\right)}
\end{equation}
where $\beta$ is a temperature to control the negative triple weight. This is a setting commonly used in KGC models.

\subsection{Collaborative Modality Adversarial Training}

\par The key function of the proposed {\fusionmodule} framework, despite achieving dual adaptive multi-modal fusion amongst imbalanced and diverse multi-modal data, is essentially feature selection for prediction, leaving the original imbalanced modality information intact. Inspired by the idea of generative adversarial networks \cite{goodfellow_generative_2014_gan, DBLP:journals/internet/Tygar11-AML}, we propose a \textbf{Collaborative Modality Adversarial Training} (abbreviated as \textbf{{\advmodule}}) module to augment the modality embeddings. The {\advmodule} module enhances the multi-modal embeddings through adversarial training, using entity-specific collaborative modality data to balance the multi-modal information distribution.

\par Drawing from the design principles of classical Wasserstein GAN (WGAN) \cite{DBLP:journals/corr/ArjovskyCB17-WGAN}, our goal is to establish a min-max game between the discriminator $\mathcal{D}$ and the generator $\mathcal{G}$ as:

\begin{equation}
    \max_{\mathcal{D}}\min_{\mathcal{G}} \mathbb{E}_{x\sim p_{real}}[{\mathcal{D}}(x)]-\mathbb{E}_{x'\sim \mathcal{G}}[{\mathcal{D}}(x')]
\end{equation}
where $x$ is the data sample. In this min-max game, the discriminator $\mathcal{D}$ is assigned to discriminate the input data sample  $x$ with a score while the generator $\mathcal{G}$ aims to generate a synthetic data sample $x'$. With the adversarial training, the generator $\mathcal{G}$ adapts to the actual distribution of the data $x$ and the discriminator $\mathcal{D}$ can learn to judge the plausibility of the input data.

\par In our scenario, the data $x$ corresponds to the entity embeddings in the KG. As mentioned previously, each entity is represented by several modal embeddings, which can denoted as
$\bm{e}_{real}=\{\bm{e}_{m_1},\bm{e}_{m_2} \dots; \bm{e}_{m_N}\}$ where $m_i \in \mathcal{M}\cup\{S\}$ is each modality. We expect to design a pair of $\mathcal{G}$ and $\mathcal{D}$ to learn the joint distribution of the multi-modal embeddings and augment the imbalanced multi-modal information. Therefore, $\mathcal{G}$ seeks to produce a collection of \textbf{synthetic embeddings}, defined as: $\bm{e}_{syn}=\{\bm{e}'_{m_1},\bm{e}'_{m_2} \dots; \bm{e}'_{m_N}\}$, and these embeddings can form a new \textbf{synthetic entity} denoted as $e^*$. By facilitating adversarial training between real and synthetic entities, we can enrich the multi-modal embeddings of the entity.

\subsubsection{\textup{\textbf{The Design of Generator}}}
\par In the design of {\advmodule}, the generator $\mathcal{G}$ is implemented by a two-layer MLP. $\mathcal{G}$ processes the input random noisy $z$ and the \textbf{concated} real multi-modal embedding $\bm{e}_{real}$ to generate augmented synthetic embeddings:

\begin{equation}
    \bm{e}_{syn}=\mathcal{G}(\bm{e}_{real}, z)=\mathbf{MLP}_2\big(\delta\left(\mathbf{MLP}_1[\bm{e}_{real}, z]\big)\right)
\end{equation}
where $\bm{e}_{real}$ is obtained by concatenating the different multi-modal embeddings collaboratively (as illustrated in Figure \ref{figure::model}) and the generated synthetic embedding $\bm{e}_{syn}$ has the same shape of $\bm{e}_{real}$. We can split $\bm{e}_{syn}$ by dimension to yield the generated embedding $\bm{e}'_{m}$ of each modality generated by $\mathcal{G}$. The ultimate goal of this design is to emulate the probability distribution of the multimodal embedding for each entity. Given a triple $(h, r, t)$, synthetic embeddings for both head and tail can be generated for both head and tail, denoted as $\bm{h}_{gen}$ and $\bm{t}_{gen}$ respectively. Besides, the corresponding synthetic entities are denoted as $h^*, t^*$.

\subsubsection{\textup{\textbf{The Design of Discriminator}}}

\par The next step is to design the discriminator $\mathcal{D}$ to scoring the synthetic embeddings. In existing works in other fields \cite{DBLP:journals/corr/ArjovskyCB17-WGAN,DBLP:conf/naacl/LiuSHWLYXC22-WGAN2}, $\mathcal{D}$ is usually implemented with another two-layer MLP. However, in the KGC task, the score function $\mathcal{F}$ mentioned in Equation \ref{score} serves as a natural alternative for the discriminator. For a given triple $(h,r,t)$ and the synthetic entities $h^*, t^*$ generated by $\mathcal{G}$, we can construct a series of synthetic triples denoted as $\{(h^*, r, t), (h, r, t^*), (h^*, r, t^*)\}$. These synthetic triples' scores can be computed with the {\fusionmodule} module and the score function $\mathcal{F}$, indicating the plausibility of the synthetic embeddings in the triple context. We can denote these synthetic triples as a \textbf{synthetic set} $\mathcal{S}(h, r, t)$. Therefore, in the design of {\advmodule}, the adversarial training loss can be denoted as:

\begin{equation}
    \mathcal{L}_{adv}=\sum_{(h, r, t)\in \mathcal{T}}\Big(-\mathcal{F}(h, r, t)+\frac{1}{|\mathcal{S}|}\sum_{(h^*,r,t^*)\atop\in\mathcal{S}(h, r, t)}\mathcal{F}(h^*,r,t^*)\Big)
\end{equation}
Note that the synthetic entities, analogous to real entities, also have several multi-modal embeddings like the real entities. They can obtain their joint embeddings with the {\fusionmodule} and calculate the final score subsequently. Therefore, the final min-max game between $\mathcal{D}$ and $\mathcal{G}$ is represented as:
\begin{equation}
    \min_{\mathcal{D}}\max_{\mathcal{G}} \mathcal{L}_{adv}
\end{equation}
According to the previous setting, the parameters of $\mathcal{G}$ encompass all the parameters in the two-layer MLP, while the parameters of $\mathcal{D}$ include the existing real embeddings, the extra parameters mentioned in {\fusionmodule}. The two parts are \textbf{iteratively} optimized with the adversarial loss to achieve convergence. 

\input{chapters/5.1-table}

\subsection{Overall Training Objective}
The final training objective of {\model} combines the above-mentioned KGC training loss $\mathcal{L}_{kgc}$ and the adversarial loss $\mathcal{L}_{adv}$ together. The discriminator $\mathcal{D}$ would minimize $\mathcal{L}_{kgc}$ and discriminate the synthetic entities generated by $\mathcal{G}$. Conversely, the generator $\mathcal{G}$ aims to generate high-score synthetic entities. Therefore, the training objective of $\mathcal{D}$ and $\mathcal{G}$ can be expressed separately as:

\begin{equation}
    \begin{cases}
        \mathcal{L}_{\mathcal{D}}=\mathcal{L}_{kgc}+\lambda_1\mathcal{L}_{adv}\\
        \mathcal{L}_{\mathcal{G}}=-\mathcal{L}_{adv}+\lambda_2 \mathcal{L}_{gp}
    \end{cases}
\end{equation}
where $\lambda_1, \lambda_2$ are the loss weights, $\mathcal{L}_{gp}$ is the gradient penalty \cite{DBLP:conf/nips/GulrajaniAADC17-GP}, commonly used for stable WGAN training, denoted as:

\begin{equation}
    \mathcal{L}_{gp}=-\sum\nolimits_{(h^*,r,t^*)\in\mathcal{S}(h^*, r, t^*)}\left(||\nabla\mathcal{F}(h^*,r,t^*)||_2-1\right)^2
\end{equation}
These two losses will be iteratively minimized during training.

\subsection{Theoretical Analysis}
In the previous sections, we give an intuitive motivation and design for the {\advmodule} module. We further provide theoretical analysis to discuss the rationale for our design.
The success of the WGAN restriction discriminator's scoring function relies on its adherence to the K-Lipschitz condition, typically accomplished through a two-layer MLP with gradient penalty \cite{DBLP:journals/corr/ArjovskyCB17-WGAN, DBLP:conf/naacl/LiuSHWLYXC22-WGAN2}. However, our design employs the RotatE score function $\mathcal{F}$ as a discriminator. We can prove that function $\mathcal{F}$ is also a K-Lipschitz function for $h$ and $t$.

\noindent\textbf{Proposition.} Let function $\mathcal{F}=-|| \bm{h}\circ\bm{r}-\bm{t}||$. Then for any $h_1,h_2$ in the function domain, there exists a scalar $K$ which satisfies:
\begin{equation}
    ||\mathcal{F}(h_1,r,t)-\mathcal{F}(h_2,r,t)||\leq K ||\bm{h}_1-\bm{h}_2||
\end{equation}

\noindent\textbf{Proof.} According to the properties of trigonometric inequalities, we have the following derivations:
\begin{equation}
    \begin{aligned}
LHS&=\big|\big| ||\bm{h}_{1}\circ\bm{r}-\bm{t}||-||\bm{h}_{2}\circ\bm{r}-\bm{t}||\big|\big|\\&\leq ||(\bm{h}_{1}\circ\bm{r}-\bm{t})-(\bm{h}_{2}\circ\bm{r}-\bm{t})||\\
&=||(\bm h_{1}-\bm h_{2})\circ r||=||\bm{r}||\times||\bm{h}_{1}- \bm{h}_{2}||
\end{aligned}
\end{equation}
Besides, in the RotatE score function, the relation embeddings $\bm{r}$ are limited with unit modulus \cite{sun_rotate_2019-RotatE} which means $||\bm{r}||\leq 1$ for any $r \in \mathcal{R}$. Let $K=1$, the proof is complete. This establishes that $\mathcal{F}$ is K-Lipschitz for head $h$. A similar argument confirms the same for the tail t. From the preceding analyses, it emerges that our WGAN-based design, harnessing the score function from the MMKGC model, is theoretically sound and rational for scoring for the synthetic entities. It makes the design of the whole framework more collaborative and natural. Indeed, {\advmodule} serves as a versatile framework that can be extendable to other embedding-based MMKGC models as well. We will go through the empirical evaluation to further demonstrate its effectiveness and generalization. 

%% file: chapters/5.1-table.tex
\begin{table*}[]
\caption{Statistical information of the five MMKGs in our WildKGC benchmark. We report the statistical information in the KGs. For each modality, we list its feature dimension and the number of entities with the corresponding modal information.}
\label{table::dataset}
\vspace{-12pt}
\resizebox{\textwidth}{!}{
\begin{tabular}{c|ccccc|cccccccccc}
\toprule
\multirow{2}{*}{\textbf{Dataset}} & \multirow{2}{*}{\textbf{\#Entity}} & \multirow{2}{*}{\textbf{\#Relation}} & \multirow{2}{*}{\textbf{\#Train}} & \multirow{2}{*}{\textbf{\#Valid}} & \multirow{2}{*}{\textbf{\#Test}} & \multicolumn{2}{c}{\textbf{Image}} & \multicolumn{2}{c}{\textbf{Text}} & \multicolumn{2}{c}{\textbf{Numeric}} & \multicolumn{2}{c}{\textbf{Audio}} & \multicolumn{2}{c}{\textbf{Video}}\\
 &  &  &  &  &  & Num & Dim & Num & Dim & Num & Dim & Num & Dim & Num & Dim\\
\midrule
\textbf{MKG-W} \cite{xu_relation-enhanced_2022-MMRNS} & 15000 & 169 & 34196 & 4276 & 4274 & 14463 & 383 & 14123 & 384 & - & - & - & - & - & -\\
\textbf{MKG-Y} \cite{xu_relation-enhanced_2022-MMRNS} & 15000 & 28 & 21310 & 2665 & 2663 & 14244 & 383 & 12305 & 384 & - & - & - & - & - & -\\
\textbf{DB15K} \cite{DBLP:conf/esws/LiuLGNOR19-MMKG} & 12842 & 279 & 79222 & 9902 & 9904 & 12818 & 4096 & 9078 & 768 & 11022 & 768 & - & - & - & -\\
\textbf{TIVA} \cite{DBLP:conf/mm/WangMCML023-TIVA} & 11858 & 16 & 20071 & 2000 & 2000 & 11636 & 2048 & 11858 & 300 & - & - & 2441 & 128 & 10269 & 2048\\
\textbf{KVC16K} \cite{DBLP:journals/corr/abs-2211-00732-KuaiPedia} & 16015 & 4 & 180190 & 22523 & 22525 & 14822 & 768 & 14822 & 768 & - & - & 14822 & 768 & 14822 & 768\\
\bottomrule
\end{tabular}
}
\vspace{-12pt}
\end{table*}

%% file: chapters/5-experiments.tex
\section{Experiments and Evaluation}

In this section, we first introduce our experimental procedure and settings, followed by a comprehensive discussion of extensive experiments to highlight the strengths of our method across a variety of scenarios. The following six research questions (RQ) are the key questions that we explore in the experiments.

\begin{itemize} % [leftmargin=16pt]
    \item[\textbf{RQ1.}] Can our model {\model} outperform the existing baseline and make substantial progress in the MMKGC task?
    \item[\textbf{RQ2.}] Can {\model} maintain robust performance in the MMKGC task when the modality information is imbalanced?
    \item[\textbf{RQ3.}] How much do each module in the {\model} contribute to the final results? Are these modules reasonably designed?
    \item[\textbf{RQ4.}] Is the {\advmodule} strategy we designed universal and general enough to be applied in other MMKGC models?
    \item[\textbf{RQ5.}] How does the training efficiency of our model compare to existing methods?
    \item[\textbf{RQ6.}] Are there intuitive cases to straightly demonstrate the effectiveness of {\model}?
\end{itemize}
\vspace{-12pt}
\input{chapters/5.1-setting}

\input{chapters/5.2-mainexp}
\input{chapters/5.3-subexp}
\input{chapters/5.4-ablation}
\input{chapters/5.5-efficency}
\input{chapters/5.6-case}

%% file: chapters/5.1-setting.tex
\subsection{Experiment Settings}
\input{chapters/5.2-table}
\subsubsection{\textup{\textbf{Datasets}}}
To better explore the MMKGC tasks in a more complex and diverse environment, we construct a new MMKGC benchmark including five different datasets. While four of the datasets are from prior studies, one is completely new. Our benchmark includes the following datasets:
\begin{itemize}[leftmargin=8pt]
    \item \textbf{MKG-Y} and \textbf{MKG-W} \cite{xu_relation-enhanced_2022-MMRNS} are two MMKGs derived from YAGO \cite{DBLP:conf/www/yago} and Wikidata \cite{DBLP:journals/cacm/wikidata} with images and texts proposed by \cite{xu_relation-enhanced_2022-MMRNS}.
    \item \textbf{DB15K} \cite{DBLP:conf/esws/LiuLGNOR19-MMKG} is an MMKG with image, text, and numerical information proposed by \cite{DBLP:conf/esws/LiuLGNOR19-MMKG}, which is a subset of DBpedia \cite{DBLP:journals/semweb/DBPedia}.
    \item \textbf{TIVA} \cite{DBLP:conf/mm/WangMCML023-TIVA} is an MMKG with image, text, audio, and video information modality with 12K entities.
    \item \textbf{KVC16K} \cite{DBLP:journals/corr/abs-2211-00732-KuaiPedia} is modified from KuaiPedia \cite{DBLP:journals/corr/abs-2211-00732-KuaiPedia}, a video concept encyclopedia. We reorganize it into an MMKG and leverage the image/text/audio/video features provided by the original authors.
\end{itemize}
We refer to our benchmark as WildKGC with the 5 MMKGs. Detailed information about the datasets is presented in Table \ref{table::dataset}. The raw modality features are inherited from the original datasets.

\subsubsection{\textup{\textbf{Task and Evaluation Protocol}}}
We conduct link prediction \cite{bordes_translating_2013-TransE} task on the five datasets, which is a significant task of KGC. We have introduced the setting of link prediction in Section \ref{section3}. Following existing works, we use rank-based metrics \cite{sun_rotate_2019-RotatE} like mean reciprocal rank (MRR) and Hit@K (K=1, 3, 10) to evaluate the results. 

Besides, the filter setting \cite{bordes_translating_2013-TransE} is applied to remove the candidate triples already existing in the training set for fair comparisons.

\subsubsection{\textup{\textbf{Baseline Methods for Comparisons}}}
In our experiments, we employ 21 different state-of-the-art KGC and MMKGC models as our baselines for a comprehensive comparison and analysis. The baselines can be divided into four categories: uni-modal (conventional) KGC methods, multi-modal KGC methods, negative sampling methods, and numeric-aware KGC methods.

\noindent \textbf{i)} \textbf{Uni-modal KGC methods.} We select 5 state-of-the-art uni-modal KGC methods including \textbf{TransE} \cite{bordes_translating_2013-TransE}, \textbf{DistMult} \cite{yang_embedding_2015-DistMult}, \textbf{ComplEx} \cite{trouillon_complex_2016-ComplEx}, \textbf{RotatE} \cite{sun_rotate_2019-RotatE}, \textbf{PairRE} \cite{chao_pairre_2021-PairRE}, which design elegant score functions and learn the structural embeddings of the given KG without any multi-modal information.

\noindent \textbf{ii)} \textbf{Multi-modal KGC methods.} We employ 10 different MMKGC methods that consider both multi-modal information and the triple structural information including \textbf{IKRL} \cite{xie_image-embodied_2017-IKRL}, \textbf{TBKGC} \cite{sergieh_multimodal_2018-TBKGC}, \textbf{TransAE} \cite{wang_multimodal_2019-TransAE}, \textbf{RSME} \cite{wang_is_2021-RSME}, \textbf{MMKRL} \cite{DBLP:journals/apin/LuWJHL22-MMKRL}, \textbf{VBKGC} \cite{DBLP:journals/corr/abs-2209-07084-VBKGC}, \textbf{OTKGE} \cite{cao_otkge_2022-OTKGE}, \textbf{IMF} \cite{li_imf_2023-IMF}, \textbf{QEB} \cite{DBLP:conf/mm/WangMCML023-TIVA} and \textbf{VISTA} \cite{lee_vista_2023-VISTA}. Comparison with these methods can verify the effectiveness of our model {\model}. Meanwhile, among these methods, \textbf{MMKRL} design an adversarial training framework. The design concepts of both methods bear some resemblance to our design of {\advmodule}.

\noindent \textbf{iii)} \textbf{Negative sampling methods.} We employ 3 different negative sampling methods to compare with our method, including \textbf{KBGAN} \cite{cai_kbgan_2018-KBGAN}, \textbf{MANS} \cite{DBLP:conf/ijcnn/ZhangCZ23-MANS}, \textbf{MMRNS} \cite{xu_relation-enhanced_2022-MMRNS}. Among these methods, \textbf{KBGAN} \cite{cai_kbgan_2018-KBGAN} is an adversarial negative sampling method designed for conventional KGC, which applies reinforcement learning to optimize the models. \textbf{MANS} \cite{DBLP:conf/ijcnn/ZhangCZ23-MANS} and \textbf{MMRNS} \cite{xu_relation-enhanced_2022-MMRNS} are two negative sampling strategies designed for MMKGC, which utilize the multi-modal information to enhance the negative sampling process. % Comparison with these methods can be used to validate the effectiveness of the {\advmodule} module design.

\noindent iv) \textbf{Numeric-aware KGC methods.} We employ 3 popular numeric-aware KGC methods including \textbf{KBLRN} \cite{DBLP:conf/uai/Garcia-DuranN18-KBLRN}, \textbf{LiteralE} \cite{DBLP:conf/semweb/KristiadiKL0F19-LiterE}, and \textbf{KGA} \cite{DBLP:conf/ijcai/WangISY22-KGA}. These methods consider the numerical information to enhance the KGC models. However, their design can not be generalized to other modalities such as image and text as they are designed only for numerical modality augmentation.

\par All of the selected baselines are embedding-based KGC and MMKGC methods. Other methods such as text-based methods \cite{DBLP:journals/corr/KG-BERT, DBLP:conf/acl/KGT5} or GNN-based methods \cite{DBLP:conf/nips/ZhuZXT21-NBFNET} are not considered as they are orthogonal to our design.

% \vspace{-16pt}
\subsubsection{\textup{\textbf{Implemention Details}}}
We implement our {\model} framework based on OpenKE \cite{DBLP:conf/emnlp/OpenKE}, a famous open-source KGC library. We conduct each experiment on a Linux server with Ubuntu 20.04.1 operating system and a single NVIDIA A800 GPU.

\par In the {\model}, we fix the batch size to 1024 and set the training epoch to 1000. The embedding dimensions $d_e, d_r$ are tuned from $\{150, 200, 250\}$ and the negative sampling number $K$ is tuned from $\{32, 64, 128\}$. The margin $\gamma$ is tuned from $\{3, 6, 9, 12\}$ and the temprature $\beta$ is set to 2. In the {\advmodule}, the random noise dimension is set to 64, and the coefficient $\lambda_1, \lambda_2$ are tuned from $\{1e^{-2}, 1e^{-3}, 1e^{-4}, 1e^{-5}\}$. We optimize the model with Adam \cite{DBLP:journals/corr/KingmaB14-Adam} and the learning rate is tuned from $\{1e^{-3}, 1e^{-4}, 1e^{-5}\}$ for both $\mathcal{G}$ and $\mathcal{D}$ in the adversarial training setting. For baselines, we reproduce the results on WildKGC following the methodology and parameter setting described in the original papers and their open-source official code. Some of the baseline results refer to MMRNS \cite{xu_relation-enhanced_2022-MMRNS}.

%% file: chapters/5.2-table.tex
\begin{table*}[]
\caption{The main MMKGC results on WildKGC benchmark. We list the modalities considered by each method where S/I/T denotes Structure/Image/Text. "All" represents that the method can process any number of input modalities.
The best results in baselines are \underline{underlined} and we highlight the SOTA results in bold. The flag $\dag$ denotes the adversarial training baselines.}

\vspace{-8pt}

\label{table::main}
\resizebox{0.85\textwidth}{!}{
\begin{tabular}{c|c|cc|cc|ccc|ccc|ccc}
\toprule
\multirow{2}{*}{Method} & \multirow{2}{*}{Modality} & \multicolumn{2}{c}{MKG-W} & \multicolumn{2}{c}{MKG-Y} & \multicolumn{3}{c}{DB15K} & \multicolumn{3}{c}{KVC16K} & \multicolumn{3}{c}{TIVA}\\
 &  & MRR & Hit@1 & MRR & Hit@1 & MRR & Hit@1 & Hit@10 & MRR & Hit@1 & Hit@10 & MRR & Hit@1 & Hit@10\\
\midrule
\multicolumn{15}{c}{\textit{\textbf{Uni-modal KGC Methods}}}\\
\midrule
\textbf{TransE} & S & 29.19 & 21.06 & 30.73 & 23.45 & 24.86 & 12.78 & 47.07 & 8.54 & 0.64 & 23.42 & 83.85 & 83.20 & 84.15\\
\textbf{DistMult} & S & 20.99 & 15.93 & 25.04 & 19.33 & 23.03 & 14.78 & 39.59 & 6.37 & 3.03 & 12.61 & 82.27 & 81.15 & 84.22\\
\textbf{ComplEx} & S & 24.93 & 19.09 & 28.71 & 22.26 & 27.48 & 18.37 & 45.37 & 12.85 & 7.48 & 23.18 & 80.67 & 77.67 & 86.10\\
\textbf{RotatE} & S & 33.67 & 26.80 & 34.95 & 29.10 & 29.28 & 17.87 & 49.66 & 14.33 & 8.25 & 26.17 & 84.59 & 83.47 & 86.95\\
\textbf{PairRE} & S & 34.40 & 28.24 & 32.01 & 25.53 & 31.13 & 21.62 & 49.30 & - & - & - & - & - & -\\
% \textbf{GC-OTE} & S & 33.92 & 26.55 & 32.95 & 26.77 & 31.85 & 22.11 & 51.18 & - & - & - & - & - & -\\
\midrule
\multicolumn{15}{c}{\textit{\textbf{Multi-modal KGC Methods}}}\\
\midrule
\textbf{IKRL} & S+I & 32.36 & 26.11 & 33.22 & 30.37 & 26.82 & 14.09 & 49.09 & 11.11 & 5.42 & 22.39 & 67.71 & 63.72 & 75.67\\
\textbf{TBKGC} & S+I+T & 31.48 & 25.31 & 33.99 & 30.47 & 28.40 & 15.61 & 49.86 & 5.39 & 0.35 & 15.52 & 81.57 & 78.75 & 86.05\\
\textbf{TransAE} & S+I & 30.00 & 21.23 & 28.10 & 25.31 & 28.09 & 21.25 & 41.17 & 10.81 & 5.31 & 21.89 & 79.57 & 74.95 & 88.07\\
\textbf{MMKRL}$\dag$ & S+I+T & 30.10 & 22.16 & 36.81 & 31.66 & 26.81 & 13.85 & 49.39 & 8.78 & 3.89 & 18.34 & 85.03 & 81.92 & 90.10\\
\textbf{RSME} & S+I & 29.23 & 23.36 & 34.44 & 31.78 & 29.76 & 24.15 & 40.29 & 12.31 & 7.14 & 22.05 & 40.01 & 30.55 & 51.35\\
\textbf{VBKGC} & S+I+T & 30.61 & 24.91 & \underline{37.04} & \underline{33.76} & 30.61 & 19.75 & 49.44 & \underline{14.66} & \underline{8.28} & \underline{27.04} & 74.07 & 66.87 & 85.85\\
\textbf{OTKGE} & S+I+T & 34.36 & 28.85 & 35.51 & 31.97 & 23.86 & 18.45 & 34.23 & 8.77 & 5.01 & 15.55 & 35.28 & 30.45 & 41.98\\
\textbf{IMF} & S+I+T & 34.50 & 28.77 & 35.79 & 32.95 & 32.25 & \underline{24.20} & 48.19 & 12.01 & 7.42 & 21.01 & 55.46 & 41.87 & 77.57\\
\textbf{QEB} & All & 32.38 & 25.47 & 34.37 & 29.49 & 28.18 & 14.82 & 51.55 & 12.06 & 5.57 & 25.01 & 74.25 & 66.10 & 88.35\\
\textbf{VISTA} & S+I+T & 32.91 & 26.12 & 30.45 & 24.87 & 30.42 & 22.49 & 45.94 & 11.89 & 6.97 & 21.27 & 76.07 & 70.67 & 86.60\\
\midrule
\multicolumn{15}{c}{\textit{\textbf{Negative Sampling Methods}}}\\
\midrule
\textbf{KBGAN} $\dag$ & S & 29.47 & 22.21 & 29.71 & 22.81 & 25.73 & 9.91 & \underline{51.93} & 13.72 & 7.54 & 25.88 & 85.44 & 82.45 & 90.10\\
\textbf{MANS} & S+I & 30.88 & 24.89 & 29.03 & 25.25 & 28.82 & 16.87 & 49.26 & 10.42 & 5.21 & 20.45 & \underline{85.70} & 82.70 & \underline{90.62} \\
\textbf{MMRNS} & S+I+T & \underline{35.03} & \underline{28.59} & 35.93 & 30.53 & \underline{32.68} & 23.01 & 51.01 & 13.31 & 7.51 & 24.68 & 83.12 & \underline{83.05} & 83.25\\
\midrule
\midrule
\textbf{{\model}} & S+I+T & 36.58 & {29.56} & {39.04} & {34.79} & 36.74 & 26.87 & \textbf{54.65} & 15.26 & 8.56 & 28.29 & 91.27 & 90.57 & 91.27\\
\textbf{{\model}} & All & \textbf{36.58} & \textbf{29.56} & \textbf{39.04} & \textbf{34.79} & \textbf{37.16} & \textbf{28.01} & 54.13 & \textbf{15.76} & \textbf{9.23} & \textbf{28.55} & \textbf{92.10} & \textbf{91.40} & \textbf{92.85}\\
\multicolumn{2}{c|}{\textbf{\textsc{Improvement}}} & +4.42\% & +3.39\% & +5.40\% & +3.05\% & +13.71\% & +15.75\% & +5.24\% & +7.50\% & +11.47\% & +5.58\% & +7.46\% & +10.05\% & +2.46\%\\
\bottomrule
\end{tabular}
}
\end{table*}

%% file: chapters/5.2-mainexp.tex
\vspace{-16pt}
\subsection{Main Results (RQ1)}

The main results of MMKGC experiments are presented in Table \ref{table::main}. We first compare our method {\model} with 18 different KGC baselines. For equitable comparisons, we account for all modalities leveraged by each method and conduct two different sets of experiments on our model. The first set considers only modalities common to all datasets (structure/image/text) and the second set considers all the modalities inherent to each specific dataset.

\par Table \ref{table::main} shows that {\model} outperforms all the existing baselines and achieves new SOTA results. Specifically, {\model} significantly surpasses the baselines by a large margin on Hit@1 on DB15K (15.75\%), KVC16K (11.47\%), and TIVA (10.05\%), indicating a marked improvement in its accurate reasoning ability. When employing only the common modalities (structure/image/text) considered by most mainstream methods, {\model} can still perform better than the baseline methods, signifying its ability to efficiently utilize information from different modalities. The performance comparison with two adversarial-based methods MMKRL \cite{DBLP:journals/apin/LuWJHL22-MMKRL} and KBGAN \cite{cai_kbgan_2018-KBGAN} demonstrates the superior effectiveness of our adversarial module design which can fully unleash the power of multi-modal information of the entities.
\input{chapters/5.2-table2}
\par Besides, none of the MMKGC baselines in Table \ref{table::main} has considered the numerical information presented in the DB15K. Accordingly, we conduct another experiment to compare our model {\model} with other numerical KGC baselines. These methods' design principles significantly differ from typical MMKGCs and lack generalizability to other modalities, so we compare them with numerical methods only on the DB15K dataset. From Table \ref{table::main2} we can observe that {\model} still outperforms all baselines when considering only the structure and numeric modalities. As the diversity of modal information increases, the effectiveness of the model is further enhanced.

%% file: chapters/5.2-table2.tex
\begin{table}[]
\caption{Comparasions among numeric-aware methods on DB15K. S/N denote the structure/numeric modalitites.}
\label{table::main2}
\vspace{-8pt}
\begin{tabular}{c|ccccc}
\toprule
Model & Modality & MRR   & Hit@1 & Hit@3 & Hit@10 \\
\midrule
\textbf{KBLN}    & S+N      & 24.68 & 19.26 & 27.27 & 34.89  \\
\textbf{LiteralE} & S+N      & 31.29 & 24.24 & 34.59 & 44.98  \\
\textbf{KGA}              & S+N      & 33.62 & 25.82 & 37.73 & 47.85  \\
\midrule
\textbf{{\model}}              & S+N      & 36.18 & 27.27 & 41.25 & 52.51  \\
\textbf{{\model}}              & All      & \textbf{37.16} & \textbf{28.01} & \textbf{42.25} & \textbf{54.13} \\
\bottomrule
\end{tabular}
\vspace{-16pt}
\end{table}

%% file: chapters/5.3-subexp.tex
\subsection{Imbalanced MMKGC Experiments (RQ2)}

\begin{figure}
  \centering
  % \vspace{-8pt}
  \subfigure[Imbalance MMKGC Experiments]{\includegraphics[width=0.9\columnwidth]{pictures/imbalance.pdf}
  }
 
  \subfigure[Group-wise Analysis for Modality-missing Triples]{\includegraphics[width=0.9\columnwidth]{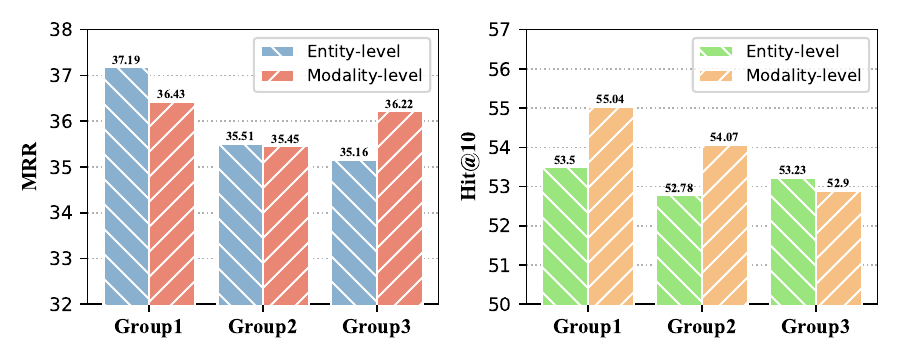}}
  \vspace{-16pt}
  \caption{The imbalance MMKGC results. We report the MRR and Hit@10 results on the DB15K datasets. Further, we divide the test triples into three groups according to whether there was complete modal information and tally their experimental results separately, where: Group1 (both h and t are modality-complete); Group2 (one of h, r is modality-missing); Group3 (both h and t are modality-missing).}
  \label{table::imbalance}
  \vspace{-16pt}
\end{figure}

\input{chapters/5.4-table}
To better illustrate the performance of {\model} in complex modality imbalance scenarios, we perform a series of MMKGC experiments in imbalance scenarios. We introduce a new parameter \textbf{imbalanced proportion} denoted as $\eta$ to quantify the imbalance in datasets, representing the percentage of entities lacking modalities information. For instance, $\eta=0.3$ means that there is 30\% missing modality information in the given MMKG.
\par Specifically, we perform a random division of the original dataset randomly selecting the corresponding proportion $\eta$ of modal information. Besides, we distinguish between two imbalance stratifications: entity-level imbalance and modality-level imbalance. For the entity-level imbalance division, we randomly drop all the modality information of a selected entity. For the modality-level imbalance, we randomly drop the modality information for a proportion $\eta$ across all entities, implying an entity might exhibit varying modal information quantities. The missing modality information dropped in these settings will be initiated randomly and the MMKGC experiments on the imbalanced datasets are shown in Figure \ref{table::imbalance}.

\par The results from the imbalanced experiments present that {\model} maintains its performance when the modality information is imbalanced. As imbalance intensifies, the performance of the baseline models {TBKGC} \cite{sergieh_multimodal_2018-TBKGC} and QBE \cite{DBLP:conf/mm/WangMCML023-TIVA} experiences a significant downturn, while our methods show relative stability. Further, the impact of imbalanced modality is more noticeable on coarse-grained metrics such as Hit@10, suggesting that the completeness of modality information has a greater influence on coarse-grained entity ranking. Additionally, a surplus of irrelevant noise within the modality information hampers the accurate inference of the model, leading to a marginal improvement in the performance of {\model} when the imbalance rate is increased. It is also noticeable that the {\advmodule} design has a significant effect on model performance, retaining its usefulness even in unbalanced situations.

\par We evaluate the imbalanced modality impact on different triples by counting separately the results of triple at different levels of missing entity modal information, which is shown in Figure \ref{figure::distribution}(b). In this figure, the proportion $\eta=0.5$. Interestingly, the model performance is not necessarily worse when modality information is fully missing (Group 3) in the triples compared to a partial missing (Group 2). % The Hit@10 of entity-level imbalance setting and MRR of modality-level imbalance both achieve the worst among their groups. Besides, there are subtle differences in model performance under different imbalance settings and different evaluation metrics.
This outcome arises because modality-level missing is much less likely to result in a complete lack of entity modal information, preserving some modality information instrumental for the final prediction, thereby leading to a relatively better performance of the model in this case (MRR of Group 3). Conversely, entity-level missing scenario deprives some entities of all modality information, causing a complete loss of valid information in the modality, leading to poor performance on coarse-grained metrics like Hit@10.

%% file: chapters/5.4-table.tex
\begin{table}[t]
\caption{The ablation study results on DB15K. We conduct three groups (G1/G2/G3) of experiments to validate the effectiveness of different modalities, the {\fusionmodule} module, and the {\advmodule} module respectively.}
\label{table::ablation}
\begin{tabular}{cl|cccc}
\toprule
\multicolumn{2}{c|}{\textbf{Setting}} & \textbf{MRR} & \textbf{Hit@1} & \textbf{Hit@10}\\
\midrule
\multirow{3}{*}{G1} & \textbf{w/o image} & 36.88 & 27.08 & 54.63\\
 & \textbf{w/o text} & 36.09 & 26.86 & 52.92\\
 & \textbf{w/o numeric} & 36.74 & 26.87 & 54.65\\
\midrule
\multirow{5}{*}{G2} & \textbf{w/o $\mathcal{P}_m$} & 35.58 & 26.12 & 52.69 \\ &
\textbf{w/o $\zeta_r$} & 35.84 & 25.79 & 53.96\\
 & \textbf{w/o $\omega_m$} & 35.65 & 25.23 & 54.10\\
 
 & \textbf{w/ concat} & 26.01 & 11.16 & 50.66\\
 & \textbf{w/ product} & 33.98 & 24.97 & 50.47\\
\midrule
\multirow{4}{*}{G3} & \textbf{w/o {\advmodule}} & 31.17 & 18.25 & 53.52\\
 & \textbf{w/o $\mathcal{L}_{gp}$} & 37.14 & 27.93 & 54.21\\
 & \textbf{w/ vanilla GAN} & 36.91 & 27.02 & 54.74\\
 & \textbf{w/ MLP as $\mathcal{D}$} & 36.34 & 27.12 & 53.31\\
\midrule
\multicolumn{2}{c|}{\textbf{Full {\model} Model}} & 37.16 & 28.01 & 54.13\\
\bottomrule
\end{tabular}
\end{table}

%% file: chapters/5.4-ablation.tex
\subsection{Ablation Study (RQ3)}
% 低质量，重改
We conduct extensive ablation studies to validate the effectiveness of each module in our design. In the ablation study, we conduct three groups of experiments to validate the contribution of each module in {\model} by removing the corresponding key component. Each of the three groups of experiments targets the different modalities, the {\fusionmodule} module, and the {\advmodule} module. The ablation study results are summarized in Table \ref{table::ablation}. From the results in G1, we can conclude that each modality contributes to the overall performance, despite the image modal seemingly having a lesser role in comparison to the text. G2 shows the essential role of key components in the {\fusionmodule} module, as their removal leads to a marked performance decline. Besides, compared with other modality fusion strategies widely used by other MMKGC models like concat and dot-product, our {\model} module still outperforms. From the results in G3, we can find that the {\advmodule} leads to a huge performance boost, especially on the precision metrics like Hit@1. Besides, when removing the gradient penalty loss or changing the adversarial loss to a vanilla GAN version with log-likelihoods \cite{goodfellow_generative_2014_gan} loss, the model performance still decreases. We also try to replace  the discriminator $\mathcal{D}$ with a two-layer MLP, but the final result is worse than the original {\model}. This suggests that the WGAN \cite{DBLP:journals/corr/ArjovskyCB17-WGAN} framework employed by us works better for the MMKGC scenario, affirming the correctness of our prior analyses and proofs in the MMKGC setting. 

%% file: chapters/5.5-efficency.tex
\subsection{Further Analysis (RQ4 \& RQ5)}
\subsubsection{\textup{\textbf{Generalization Analysis (RQ4)}}}
\begin{figure}[]
  \centering
\includegraphics[width=0.9\linewidth]{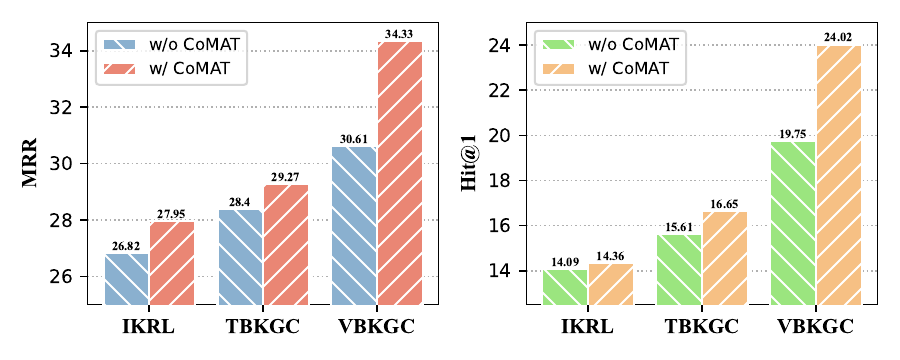}
  \vspace{-16pt}
  \caption{The generalization experiments of the {\advmodule} module on three different MMKGC models. We report the MRR and Hit@1 results on the DB15K dataset.}
  \label{figure::generalize}
  \vspace{-8pt}
\end{figure}

As we mentioned before, {\advmodule} is a general framework for enhancing MMKGC models based on embedding. To prove this point, we apply {\advmodule} on more MMKGC models (IKRL \cite{xie_image-embodied_2017-IKRL}, TBKGC \cite{sergieh_multimodal_2018-TBKGC}, and VBKGC \cite{DBLP:journals/corr/abs-2209-07084-VBKGC}) and demonstrate the results in Figure \ref{figure::generalize}. 

\par We can make a conclusion that the MMKGC models trained w/ {\advmodule} can obtain significant performance gain. This suggests that {\advmodule} can be used as a general adversarial enhancement framework in different MMKGC models.

\begin{figure}[h]
  \centering
\includegraphics[width=\linewidth]{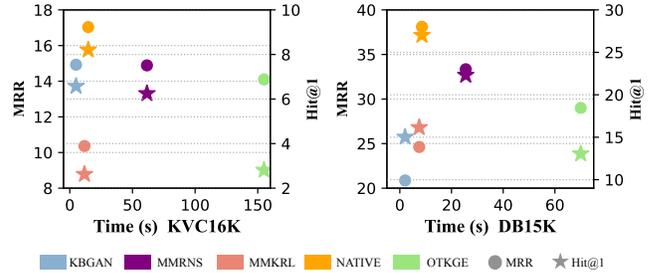}
  \vspace{-24pt}
  \caption{The results of the efficiency experiment. We report the MRR and Hit@1 results on the KVC16K/DB15K datasets.}
  % 缺图例
  \label{figure::efficient}
  \vspace{-16pt}
\end{figure}

\subsubsection{\textup{\textbf{Efficency Analysis (RQ5)}}}
Another important concern is the efficiency of the adversarial methods. This is because the iterative training strategies would lead to more computation. Therefore, we evaluate the efficiency of several MMKGC methods. The methods we have chosen cover both non-adversarial (OTKGE \cite{cao_otkge_2022-OTKGE}, MMRNS \cite{xu_relation-enhanced_2022-MMRNS}, IMF \cite{li_imf_2023-IMF}) and adversarial approaches (KBGAN \cite{cai_kbgan_2018-KBGAN}, MMKRL \cite{DBLP:journals/apin/LuWJHL22-MMKRL}). We evaluate the efficiency and performance of different models with the same batch size of 1024 and dimension of 250 on the same device.
As shown in Figure \ref{figure::efficient}, we can observe that {\model} makes a good trade-off between efficiency and performance, achieving the best results while keeping relatively fast and efficient. Though the adversarial training module {\advmodule} slows down the training to some extent, this latency is within acceptable limits and there are significant gains in the model performance.

%% file: chapters/5.6-case.tex
\subsection{Case Study (RQ6)}

\begin{figure}[]
  \centering
\includegraphics[width=0.92\linewidth]{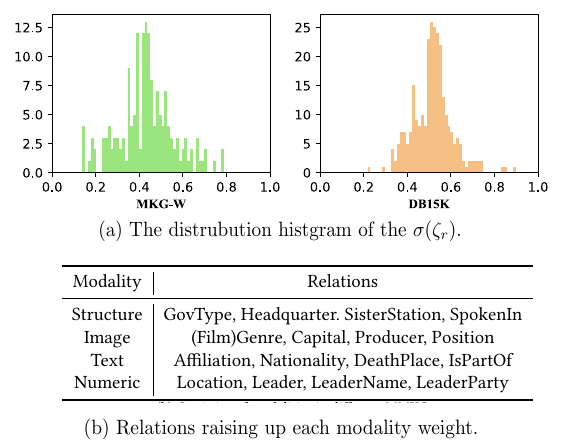}
  \vspace{-8pt}
  \caption{The relation-wise temperature distribution and high-frequency relations that can raise each modality weight.}
  \label{figure::case}
  \vspace{-12pt}
\end{figure}

This section presents clear examples to further elucidate our design. We previously emphasize that the relation guidance in {\fusionmodule} can zoom in on the differences between the different modalities and sift through them to find the useful parts, as the relation-wise temperatures are limited in $(0, 1)$ with a sigmoid function. We first demonstrate the distribution of the temperatures as shown in Figure \ref{figure::case}. We can observe that the distribution is diversified among relations. To better illustrate this, we filter out and list the most frequent relations that pull the weights of the corresponding modality higher for each modality, which can be found in Figure \ref{figure::case} (b). We can find that each different relationship will have a different dependence on the different modal information of the entity when making predictions. For example, image information can provide some intuitive visual information. When predicting movie \textit{genre}, such image information would greatly benefit the result. Meanwhile, the \textit{affiliation}, \textit{nationality}, and \textit{death place} of a person can be usually found in the text descriptions which makes text modality valuable when predicting the results of these relations.

%% file: chapters/6-conclusion.tex
\section{Conclusion}
In this paper, we propose a novel MMKGC framework {\model} to rectify the diversity and imbalance issues of the MMKGs. {\model} consists of two core designs called {\fusionmodule} and {\advmodule}. {\fusionmodule} proposes a relation-guided dual adaptive fusion method to incorporate adaptive features of any modalities to obtain joint representations while {\advmodule} aims to enhance the information about the imbalanced modes by a training strategy based on WGAN. We perform an in-depth theoretical analysis to justify our design's rationale. We construct a new MMKGC benchmark and conduct comprehensive experiments on it against 21 baselines to show the effectiveness, generalization, and efficiency of our framework. In the future, we think the MMKGC task and the downstream application of MMKG in more complex real scenarios remain great challenges to be solved.

% polish and Improve the following latex writing to make it more natural and concise, ignore the bib citation in the text: